\documentclass[10pt,journal,compsoc]{IEEEtran}
 
\makeatletter

\usepackage{graphicx}
\usepackage{relsize}

\usepackage{ragged2e}

\usepackage{subfig}

\usepackage{lipsum}

\newcommand\notsotiny{\@setfontsize\notsotiny\@vipt\@viipt}

\usepackage{caption}
\usepackage{xcolor}
\usepackage{amsmath}

\newcommand{\highlight}[1]{\textcolor{black}{#1}}
\newcommand{\newrevision}[1]{\textcolor{black}{#1}}

\usepackage[normalem]{ulem}

\usepackage[utf8]{inputenc}
\usepackage[T1]{fontenc}
\usepackage{lmodern}
 
\mathchardef\mhyphen="2D 

\newcommand{\ignore}[1]{}

\newcommand\Mark[1]{\textsuperscript{#1}}

\usepackage{hyperref}
\hypersetup{
    colorlinks=true,
    linkcolor=blue,
    filecolor=magenta,      
    urlcolor=cyan,
}

\IEEEaftertitletext{\vspace{-3\baselineskip}}

\IEEEoverridecommandlockouts

\begin{document}

\IEEEspecialpapernotice{\vspace{-24pt}}  

\title{The Bitlet Model \\ \vspace{-15pt}{\normalsize \textit{\newrevision{Defining a Litmus Test for the Bitwise Processing-in-Memory Paradigm}}} \\[.75ex] 
  {\normalfont\large 
    Kunal Korgaonkar\Mark{\$,*}, Ronny Ronen\Mark{\$}, Anupam Chattopadhyay\Mark{\#}, Shahar Kvatinsky\Mark{\$} %
  }\\[-1.5ex]
}

\author{
    \IEEEauthorblockA{%
        \Mark{\$}Technion (Israel) ~~~~ %
    }
    \and
    \IEEEauthorblockA{%
        \Mark{*}UC San Diego (US) ~~~~ %
    }
    \and
    \IEEEauthorblockA{%
        \Mark{\#}NTU (Singapore)%
    }
    \IEEEcompsocitemizethanks{\IEEEcompsocthanksitem \scriptsize{{A litmus test is a definite test which produces a decisive result.}} \IEEEcompsocthanksitem{\scriptsize $^{*}$Part of the work done while Kunal Korgaonkar was a student at UC San Diego (US).}} 
}


\ignore{  
\author{
 Kunal Korgaonkar$^{\$,@}$, 
 \and Ronny Ronen$^{\$}$, 
 \and Anupam Chattopadhyay$^{\#}$, 
 \and and Shahar Kvatinsky$^{\$}$    
}
}

\ignore{ 
\author{\vspace{-10pt} Kunal Korgaonkar, Ronny Ronen, Anupam Chattopadhyay, and Shahar Kvatinsky} 
  
\IEEEcompsocthanksitem{\scriptsize Kunal Korgaonkar, Ronny Ronen and Shahar Kvatinsky are affiliated to Technion (Israel) and Anupam Chattopadhyay is with NTU (Singapore). Part of the work was done while Kunal was a student at UC San Diego (USA). 
}

\ignore{ This work was supported by the ERC through the European Union’s Horizon 2020 Research and Innovation Programme under Grant 757259 and by the ISF under Grant 1514/17.
}

}

\IEEEtitleabstractindextext{
\justify 
\begin{abstract}
This paper describes an analytical modeling tool called Bitlet that can be used, in a parameterized fashion, to understand the affinity of workloads to processing-in-memory (PIM) as opposed to traditional computing. The tool uncovers interesting trade-offs between operation complexity (cycles required to perform an operation through PIM) and other key parameters, such as system memory bandwidth, data transfer size, the extent of data alignment, and effective memory capacity involved in PIM computations. Despite its simplicity, the model has already proven useful. In the future, we intend to extend and refine Bitlet to further increase its utility.  
\end{abstract}

\vspace*{-3mm}


\vspace*{-1mm} 

}

\maketitle

\vspace{-10mm}

\section{Introduction}
\label{sec:intro}

Processing huge amounts of data on traditional von Neumann architectures, involves many data transfers between the CPU and the memory. These transfers degrade performance and consume energy~\cite{Ranganathan2011,  Seshadri2014, Seshadri2017, Fujiki2018, Eckert2018}. Enabled by emerging memory technologies, recent processing-in-memory (PIM) solutions show great potential in reducing costly data transfers by performing computations using individual memory cells~\cite{Raoux2008, Borghetti2010, Wong2012, Linn2012, Kvatinsky2014_1}. This line of research has led to better circuits and micro-architectures~\cite{Kvatinsky2014_1, Kvatinsky2014_2, Bhattacharjee2017}, as well as applications using this paradigm~\cite{Imani2017, Haj2018}. 


Despite the recent resurgence of PIM, it is still very challenging to analyze and quantify the advantages or disadvantages of PIM solutions over other computing paradigms. We believe a useful analytical modeling tool for PIM can play a crucial role.  An analytical tool in this context has many potential uses, such as in (i) evaluation of applications mapped to PIM, (ii) comparison of PIM versus traditional architectures, and (iii) analysis of the implications of new memory technology trends on PIM. 

\ignore{ 
\highlight{For instance, current heuristic-based approaches only map simple operations (\textit{e.g.}, logical bit-wise) to PIM and may fail to capture its full capabilities. At the same time, mapping complex operations (\textit{e.g.}, divide) without a clear view of PIM limits may lead to lower-than-expected performance.}
} 

Our Bitlet model is an analytical modeling tool that addresses the challenge of better understanding PIM relative to traditional CPU/GPU computing. The name Bitlet reflects PIM's unique bit-by-bit data element processing approach. The model is inspired by past successful analytical models for computing~\cite{Gustafson1988, Hill2008, Williams2009, Esmaeilzadeh2011, Hill2019} and provides a simple operational view of PIM computations.  

\ignore{ The modeling abstracts PIM implementation through model parameters related to algorithms (such as complexity of operations), technology (such as cycle time of in-memory operations) and architecture (such as memory bandwidth). 
} 

The main contributions of this work are:  
\begin{itemize} 
\item Presentation of the Bitlet model, an analytical modeling tool that abstracts algorithmic, technological as well as architectural machine parameters for PIM.
\item 
\newrevision{Definition of a \textit{litmus test} for workloads to assess their affinity on PIM as compared to the CPU.}
\item Delineation of the strengths and weaknesses of the new PIM paradigm as observed in a sensitivity study evaluating PIM performance and efficiency over various Bitlet model parameters. 
\end{itemize}


\section{The Bitlet Model}
\label{sec:model}

We derive a parameterized throughput metric for PIM followed by one for the CPU. The throughput focus is in alignment with the parallelism that the PIM approach offers. \highlight{We first describe the model and then proceed to explain how to apply it.} Throughout the paper, we refer to `PIM' as a framework for processing inside memories. 

\subsection{Deriving PIM Throughput} 
\label{sec:pim} 

\highlight{We based the PIM side of the Bitlet model on the principle of performing computations using memristive memory arrays, wherein processing occurs inside the memory arrays using a stateful in-memory logic family (\textit{e.g.}, IMPLY~\cite{Borghetti2010} and MAGIC~\cite{Kvatinsky2014_1}). The execution does not necessitate moving data out of the memory arrays if the data is present there already. The other key principle of the proposed PIM model is its reliance on a series of simple operations to compute any complex operation inside the memories (\textit{e.g.}, MAGIC uses simple NOR as the basic operation).} 

\highlight{These principles are the foundation of what are currently known as \textit{true PIM} solutions, which offer advantages such as simplified peripheral circuitry, less reliance on additional external arithmetic units, and lower energy consumption. We base the Bitlet model on true PIM solutions, given their wide applicability and advantages. Although we use MAGIC~\cite{Kvatinsky2014_1} as an example of a stateful in-memory logic family to illustrate true PIM, our model is also easily extendable to other stateful in-memory logic families. The supporting circuitry and micro-architecture for our PIM model resemble, but are not limited to, those described by Haj-Ali \textit{et al.}~\cite{Haj2018_2}.}

\newrevision{We derive PIM throughput by considering operation complexity, data placement and alignment issues, and energy efficiency. We start by discussing operation complexity.}

\textbf{Operation Complexity.} In the Bitlet model, the PIM computations are carried out as a series of NOR operations, applied on the memory cells of a row inside a memristive memory array. Each row of the memory array stores the input data required for processing. A two-input bit NOR gate processes two data bits within the row and stores the output bit in the same row. Any intermediate data are processed similarly. Processing proceeds sequentially in this fashion to produce the final output, which is also stored within the same row. Data processing as per the Bitlet model is best viewed as \emph{row-wise} and \emph{bit-by-bit} within the row of a memory array. We use a default two-input bit NOR gate as the basic logic operation~\cite{Kvatinsky2014_1}, permitting a maximum of two input bits to be processed per memory cycle.
 
 While each row is processed bit-by-bit, the effective throughput of PIM is increased by the inherent parallelism achieved by simultaneous processing of multiple rows inside a memory array and of multiple memory arrays in the system memory. We assume the same computations (\textit{i.e.}, individual operations) applied to a row are also applied in parallel in every cycle across all the rows ($ROW$) of a memory array. This parallelism is made possible by the 2D structure of the memory arrays and by reuse of the voltage signals used to operate an individual row for all the rows. Although the choice to only process row-wise may seem restrictive, it naturally maximizes the data-level parallelism and hence PIM throughput. Moreover, the multiple memory arrays ($MAT$) further maximize this parallelism. Finally, the cycle time, $CT$, of a single basic PIM operation also impacts overall PIM performance. The shorter it is, the faster the processing.

 \begin{figure*}[t]
 \begin{center}
  \includegraphics[width=4in]{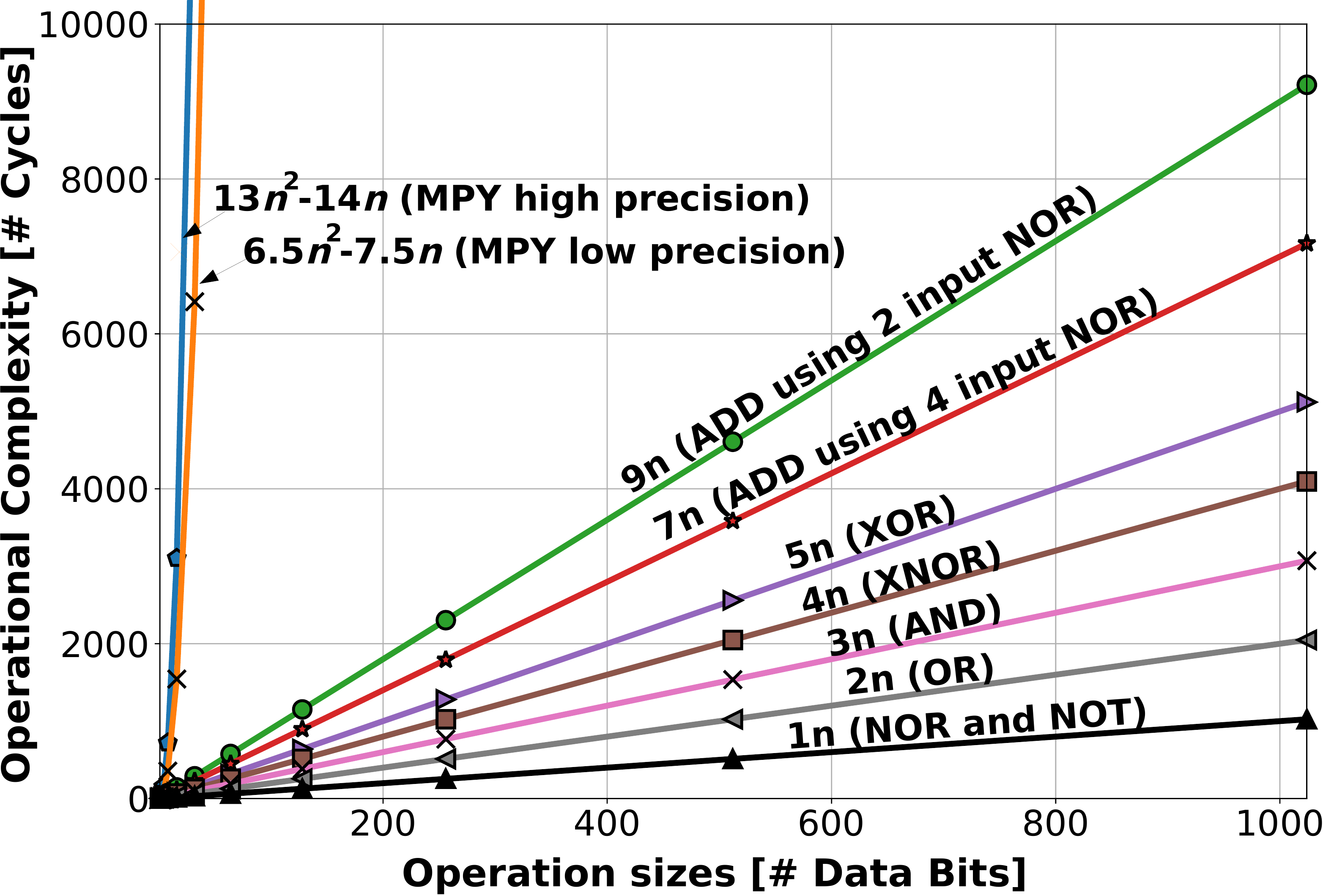}
 \caption{{PIM operation complexity in cycles for different types of operations and data sizes. MPY refers to a multiplication operation. Other arithmetic and logic operations are also shown.}}
  \label{fig:funcs}
  \end{center} 
\end{figure*}

 Fig.~\ref{fig:funcs} shows how bit lengths ($n$) of the input data affect the number of computing cycles required for PIM-based processing. The figure shows that this number is affected by both the data sizes, as well as operation types (different operations follow a different curve on the graph). With this model, for example, n-bit AND requires $3n$ cycles (\textit{e.g.}, for n=16 bits AND takes 16x3 = 48 cycles), ADD requires $9n$ cycles\footnote{\scriptsize ADD can be improved to $7n$ cycles using an algorithmic optimization that uses four-input NOR instead of two-input NOR.}, and multiply (MPY) requires 13${n}^{2}-$14$n$ cycles~\cite{Haj2018}. We define the \textbf{operation complexity} parameter ($OC$) for a given operation type and data size, as the number of cycles required to process the corresponding data.

The throughput of PIM is captured by four parameters: $OC$, $MAT$, $ROW$ and $CT$ (see Table~\ref{tab:parameters}). The throughput of the system in operations per second can be expressed as:

\begin{gather} 
 {\bf  \scriptstyle  Perf{\text -}PIM(Op)  =   \frac{ ROW x MAT}{ OC \times CT }. }\label{eq:1}  
\end{gather}


\newrevision{\textbf{Placement and Alignment Complexity.} PIM imposes certain constraints on data alignment and placement~\cite{Talati2018}. To align the data for subsequent row-parallel operations, a series of data alignment and placement steps may be needed. The number of cycles needed to perform these additional steps is captured by the placement and alignment complexity parameter, denoted as $PAC$. Currently, for simplicity, we focus on modeling the cost of intra-array data movements and assume that multiple memory arrays continue to operate in parallel and independently. We have already observed that PIM performance being quite sensitive to intra-array data movements (Section~\ref{sec:apply}). In the future, we plan to refine the model for inter-array data movements.} 

\newrevision{The following expression, extends Eq.~\ref{eq:1}, considering presence of unaligned and misplaced data elements:}   

\begin{gather} 
 {\bf \newrevision{ \scriptstyle  Perf{\text -}PIM(Op)  =   \frac{ ROW x MAT}{ (OC + PAC)  \times CT  }  .} }  
 \label{eq:2} 
\end{gather}   


\newrevision{The PAC cycles can, in turn, be broken down into a series of vertical, column-parallel moves and horizontal, row-parallel moves to bring the data in a memory array to the desired locations. While the vertical moves serve to correct the data element misplacements, the horizontal moves take care of unaligned data elements. Given, $\scriptstyle{HMOVE}$ and $\scriptstyle{VMOVE}$ as the total number of horizontal and vertical moves needed, respectively, PAC can be said to be equal to `$ \scriptstyle{(HMOVE} + \scriptstyle{{VMOVE})}$'. The horizontal moves are performed bit-by-bit for a given data element, and hence, their count is typically proportional to the size of the data element involved. In most cases, the same alignment is done for all data elements; thus, the same bit moves in parallel in all rows. When the involved data elements across different rows are not aligned, separate horizontal moves need to be made individually for each data element  (increasing the cost). A vertical move for a given data element, on the other hand, is parallelizable. However, to cover the many data elements distributed across the rows, many such vertical moves need to be performed serially.}

 \begin{table*}[!t]
 \scriptsize   
\centering 
\begin{tabular}{ |c | c | c | c |}
 \hline 
 {\bf Parameter name} &  {\bf Notation} & {\bf Value(s)} & {\bf Type} \\ \hline \hline 
 {PIM operation complexity} & $OC$ & 1 - 32k cycles & Algo.    \\ \hline  
  \newrevision{PIM Placement and } &     &    &      \\  
  \newrevision{Alignment Complexity} &  \newrevision{$PAC$}  &   \newrevision{0 - 1024x1024} & \newrevision{Algo.} \\ \hline  
 {PIM cycle time} &  $CT$ & 10 ns~\cite{Mario2019} & Tech. \\  \hline 
 {PIM array dimensions} &  \textbf{\tiny $ROW{\times}COL$} & 1024 x 1024   & Tech.      \\ \hline 
 {PIM array count } &  $MAT$  &   1k - 16k & Arch.    \\ \hline 
 PIM energy for op ($OC$=1) & $E^{PIM}$ & 0.1pJ~\cite{Mario2019}& Tech. \\ \hline \hline
CPU memory bandwidth & $BW$ & 1 to 16 Tbps & Arch. \\ \hline  
  CPU data in-out bits & $DIO$ & 24, 48 & Algo.  \\ \hline
   CPU energy for bit transfer  & $E^{CPU}$ & 15pJ~\cite{Connor2017} & Tech. \\ \hline 
\end{tabular}
\caption{\small \textit{Bitlet model parameters.}}
\label{tab:parameters} 
\end{table*}

\newrevision{As an example, if $a$, $b$ and $c$ are three data elements vectors inside a memory array and the computation requires performing `a(i) = b(i+1) + c(i)', then $b$ in this case is unaligned and also misplaced. For this scenario, each b(i+1) is relocated to t(i) through multiple horizontal moves and a single vertical move. Only after the relocations, the actual computation, which in this case is a(i)=t(i)+c(i), is performed. To relocate b(i+1) to t(i), firstly, $n$ horizontal moves occur, which ensures alignment of all b(i+1), each of size $n$, followed by as many as row count number\footnote{Here, (ROW-1) moves occur within and 1 out of the MAT.} of vertical moves, which takes care of the misplacement of each individual b(i+1) inside each row. Therefore, in this scenario, the PAC is $\scriptstyle{(n + ROW)}$ cycles.   }

 \textbf{Energy Efficiency.} The maximum throughput for PIM or the CPU is limited by the thermal design power (TDP). For PIM, the throughput depends on the energy per unit of computation, which is the energy spent for a single computation cycle ($E^{PIM}$) for $OC$ = 1. Building on Eq.~\ref{eq:2}, we quantify the power-limited (PL) throughput as follows:

\begin{equation}  {\bf \scriptstyle  
 PL{\text -}Perf{\text -}PIM(Op)= Min ( Perf{\text -}PIM(Op),\frac{ TDP}{  E^{PIM} \times (OC + \newrevision{PAC}) } ).
 } \label{eq:3} 
 \end{equation} 

Table~\ref{tab:parameters} summarizes the PIM-related parameters of the Bitlet model. For conceptual clarity and to aid our analysis, we designate three parameter types: \textit{technological}, \textit{architectural}, and \textit{algorithmic}. \highlight{Typical values, or the ranges for the different parameters, are also listed in the table.}   
\subsection{Deriving CPU Throughput} 
\label{sec:cpu} 

\newrevision{Given the objective of the Bitlet model to assess the affinity of workloads or workload phases to PIM versus CPU, the model focuses on workloads (or workload phases) with high memory intensity and relies on a relatively simple CPU model. The overall distinction in modeling between PIM and CPU is described below.}

\newrevision{For the workload phases being considered for the Bitlet litmus test, PIM-based computations (as outlined in Section~\ref{sec:pim}) occur inside the memory arrays, without any data transfers occurring outside the memory arrays. That is, they are limited by operation complexity and by the data placement and alignment costs. On the other hand, we assume that the CPU throughput is primarily limited by its usage of external memory-bandwidth, i.e., by the cost of data transfers between the CPU and memory, ignoring the cost of computations and data movements performed within the CPU itself.}

\highlight{\textbf{Data Transfer.} The Bitlet model, therefore, derives the CPU throughput assuming both memory bandwidth between the CPU and the memories, and the amount of data transfer needed to perform an operation, as the primary limiting factors.} Large amounts of data being transferred between the CPU and the memory result in lower CPU throughput, while smaller volumes produce the opposite effect. The extent of data transfer between the CPU and the memory is captured by the \textbf{data in-out} ($DIO$) model parameter. The $DIO$ is the average amount of data transferred per operation and must account for all the data transfers (in bits) between the CPU and the memory resulting from inputs, outputs, as well as any temporary results. Along with $DIO$, the external memory bandwidth (denoted as $BW$) between the CPU and the memory determines the final throughput\footnote{\scriptsize Memory bandwidth may depend on the number of channels}. The CPU throughput, in operations per second, is defined as:   

\begin{equation} {\bf \scriptstyle    Perf{\text -}CPU(Op) =  \frac{ BW }{ DIO }}. 
\label{eq:4} 
\end{equation}  

\newrevision{To support a broader analysis across all types of workloads, including the phases with high CPU arithmetic intensity, a more accurate CPU model will be useful. One possibility is the inclusion of maximum arithmetic throughput as part of Eq.~\ref{eq:4}, similar to the arithmetic intensity limit of Roofline~\cite{Williams2009}. We leave extension of the Bitlet model with more detailed CPU-side modeling, for future work.} 
  
\textbf{Energy Efficiency.} On the CPU front, the energy per bit transfer between the CPU and the memory determines the efficiency of CPU computations (denoted as $E^{CPU}$ and also listed in Table~\ref{tab:parameters}). We assume that the CPU compute energy is significantly lower than the data transfer energy. This aligns with our focus on identifying the strengths of PIM rather than those of the CPU. The power-limited performance for CPU computation is expressed as:   
\begin{equation} {\bf \scriptstyle 
PL{\text -}Perf{\text -}CPU(Op) =  Min ( Perf{\text -}CPU(Op), \frac{ TDP }{  E^{CPU} \times DIO }  ).
} 
\end{equation}

Table~\ref{tab:parameters} summarizes the CPU-related parameters, including typical values or range of values they are set to. We vary the memory bandwidth parameter from 1 to 16 Tbps to show the sensitivity of the model to memory bandwidth.

\begin{figure*}[t]
\centering
\includegraphics[width=6in]{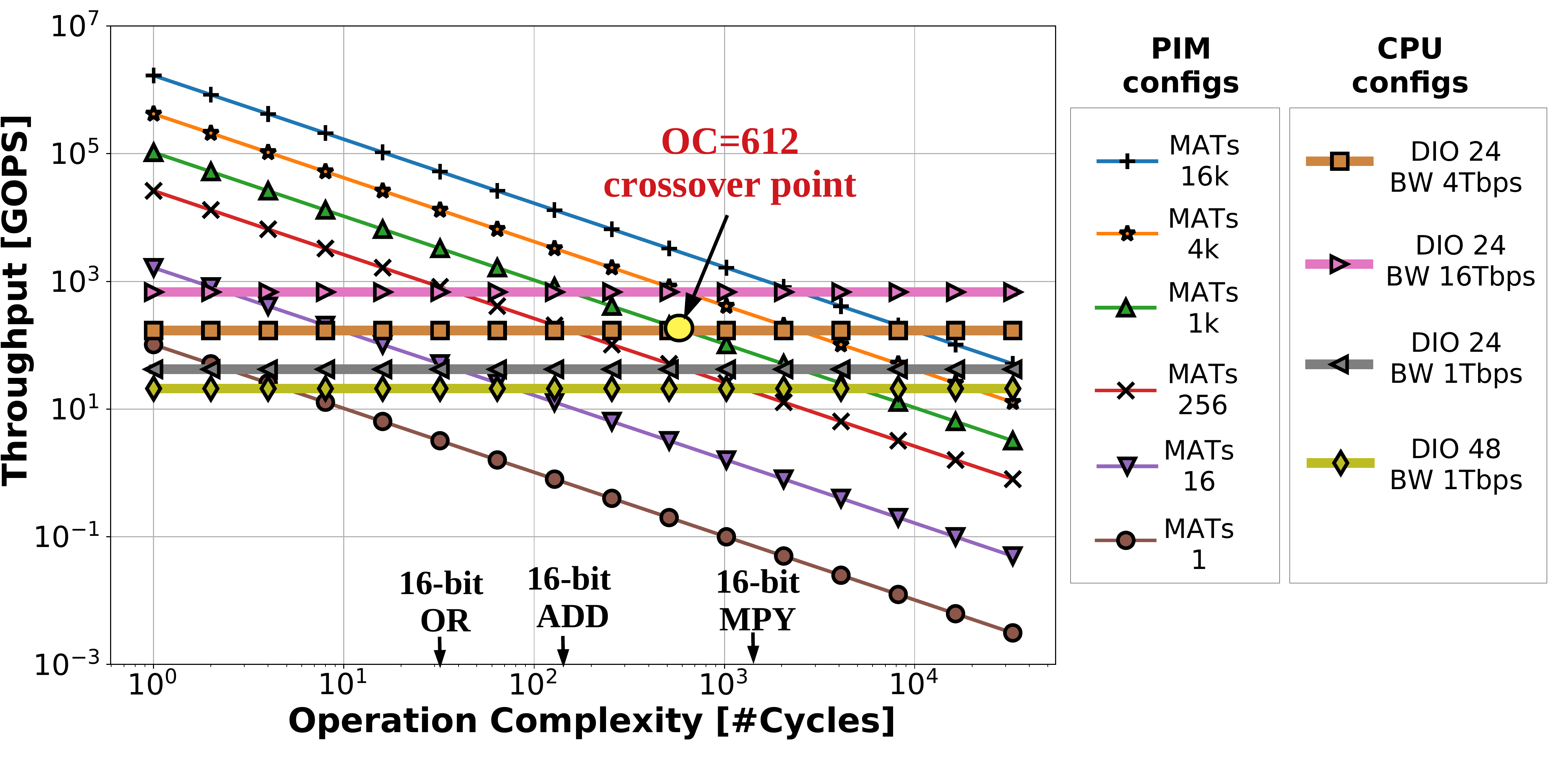}
\caption{\scriptsize {Throughput comparison of CPU vs. PIM. A crossover point where the CPU starts performing better than PIM is shown.}}
\label{fig:homo}
\end{figure*}
 
\section{{Applying the Bitlet Model}}
\label{sec:apply} 

\highlight{In this section, we apply the Bitlet model. We start by comparing the throughput of basic operations for PIM versus the CPU and then proceed to compare PIM to CPU under a wider parameter design space.}


\subsection{{PIM vs. CPU - Basic Operations}} 
\label{sec:oc1}
\highlight{Below we discuss a few examples to illustrate the use of the Bitlet model. Note that although we only compare PIM to the CPU, the model assumptions and the comparisons can be easily extended to GPUs as well.} 

\textbf{PIM (16-bit) ADD, OR and MPY.} Consider an ADD operation which adds two 16-bit inputs and produces a 16-bit output \newrevision{and assuming all data elements are perfectly aligned}. This operation on a data element takes 144 cycles ($OC$ = 144, 9$n$ where $n$=16). Assuming there are 1024 MATs and each MAT supports 1024 data elements (rows= \# data elements), the achieved throughput = (1024x1024)/(144x10) = {728} GOPS. Now consider a 16-bit OR operation, that has two 16-bit inputs and produces a 16-bit output. In this case $OC$ = 32 (2$n$, where $n$=16) and the throughput = (1024x1024)/(32x10) = {3276} GOPS. Finally, consider a 16-bit MPY (multiplication) producing a 32-bit result. In this case, $OC$ = 3104 (13${n}^{2}$- 14$n$, where $n$=16). Here, the throughput is (1024x1024)/(3104x10) = {33} GOPS. For low-precision multiplication that produces only a 16-bit output, OC = 1544 and the throughput is (1024x1024)/(1544x10) = {67} GOPS.

\textbf{CPU (16-bit) ANY.}  We consider `any' binary operation that operates on two 16-bit inputs and produces 16-bit output (\textit{e.g.} 16-bit ADD, 16-bit OR and 16-bit MPY with low-precision). The DIO is thus (16x2+16) = 48 bits\footnote{\scriptsize DIO = 24, for two 8-bit inputs and one 8-bit output.}. For any of these operations, the effective throughput of the CPU is 4Tbps/48 = {85} GOPS. For an OR operation, the CPU is inferior to PIM, which benefits, in this scenario, from lower operation complexity, high data parallelism, and obliviousness to external memory bandwidth. For MPY, on the other hand, PIM is inferior to the CPU due to the higher operation complexity. If the memory bandwidth is reduced to 1024 Gbps, the CPU throughput becomes 1Tbps/48 = {21} GOPS for any 16-bit binary operation, with a 16-bit output. Since memory bandwidth is the main limiter here, CPU throughput becomes worse than PIM even for MPY.


\subsection{{PIM vs. CPU - Impact of Model Parameters}} 
\label{sec:oc2}

\newrevision{PIM's throughput is sensitive to various Bitlet model parameters. In this section, a sensitivity study performed to assess these model parameters, highlights some of the strengths and weaknesses of the new PIM paradigm.}


\textbf{Operational Complexity Impact.} Fig.~\ref{fig:homo} shows the throughput of PIM versus that of the CPU \newrevision{and assumes PAC = 0}. Diagonal lines represent PIM with varying numbers of $MATs$ (set to 1/16/256/1024/4096/16384 $MATs$). A single 1024$\times$1024 memory array has a 128 KB capacity. Horizontal lines are for CPUs with varying $DIO$ bits (set to 24/48) along with \highlight{$BW$ = 1Tbps/4Tbps/16Tbps}.

Using Eq.~\ref{eq:1}, we observe that PIM throughput increases with maximum MAT availability, peaking when maximum available memory arrays are used ($MATs$ = 16k), and the operation complexity is the lowest possible ($OC$ = 1). In parallel, PIM throughput decreases with increasing operation complexity. We see that the CPU throughput decreases with higher $DIO$. For instance, consider the lines shown for $DIO$ = 24 and $DIO$ = 48 for the same $BW$ = 1Tbps. The CPU's performance for $DIO$ = 48 is lower than for $DIO$ = 24.  

For a configuration of $MAT$ = 1024, $DIO$ = 24 and $BW$ = 4Tbps, the CPU performs better than PIM at $OC$ = \textbf{612} or higher. This marks the crossover point and sets the boundaries of a favorable region for PIM for this configuration. Note the placement of the OR, AND and MPY operations shown in Fig.~\ref{fig:homo} along the x-axis. Clearly, OR ($OC$ = 32) and ADD ($OC$ = 144) are located to the left of the crossover point and MPY ($OC$ = 3104) is to the right. The left region is where PIM is superior, and the right region is where CPU is superior.    
 
The crossover point shifts to the right for different $DIO$ values. For instance, for $MAT$ = 1024 and $BW$ = 1024, the crossover point shifts roughly from $OC$ = 2500 to $OC$ = 5000 for $DIO$ = 24 to $DIO$ = 48, respectively. Thus, it is the algorithmic interplay of $OC$ and $DIO$ (along with other technological and architectural factors) that determines the throughput of PIM relative to that of CPU computing.   

\begin{figure*}[t]
\centering
\includegraphics[width=6in]{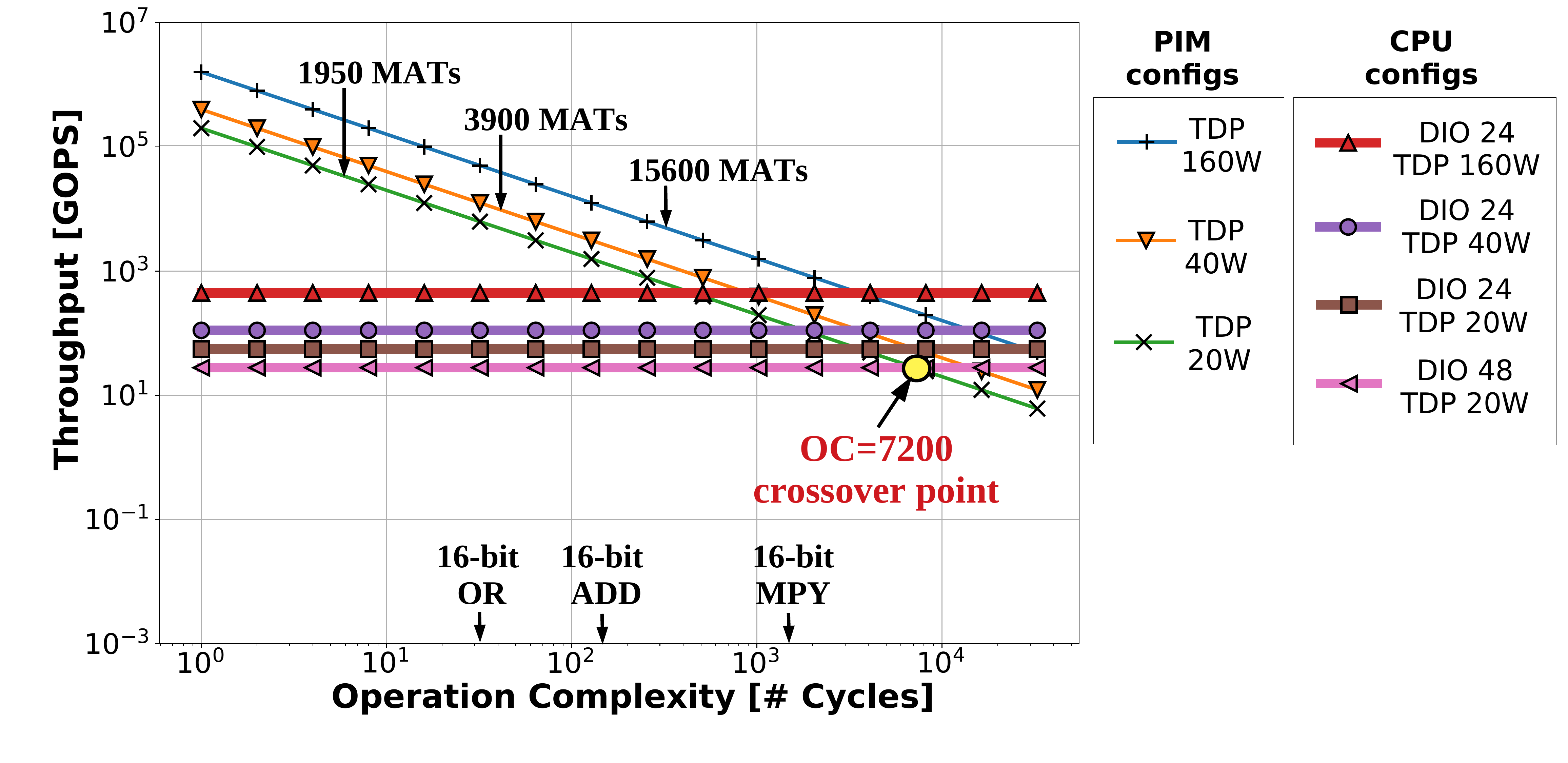}
\caption{\scriptsize {Throughput comparison of CPU vs. PIM under power limits. Shows no. of PIM MATs permissible under a power limit.}}
\label{fig:power}
\end{figure*}


\newrevision{\textbf{Placement and Alignment Complexity Impact.} Under perfect data layout, $PAC$ = 0, and therefore there is no PIM performance loss due to placement and alignment issues. Sometimes, however, either horizontal or vertical moves, or both, become necessary for placement and alignment reasons, which result in some performance loss. Below we discuss the nature and the magnitude of these losses, captured through the $PAC$ parameter and Eq.~\ref{eq:2}.}  

\newrevision{Consider the previous example `a(i)=b(i+1)+c(i)' (assuming 16-bit addition) which necessitates both horizontal and vertical moves. In this example, $PAC$ = 16+1024 = 1040 and following Eq.~\ref{eq:2}, the throughput is (1024x1024)/((144+1040)x10) = 88 GOPS. Effectively the throughput shrinks to 12\% of its original value of 728 GOPS (as per Eq.~\ref{eq:1}). Now consider a scenario requiring just the horizontal moves for alignment. In this case, prior to the additions, 16 horizontal moves are needed, \textit{i.e.,} $PAC$ = 16. In this case, the throughput is 655 GOPS, just a 10\% performance loss relative to 728 GOPS. Finally, if a subset of data elements has to be aligned separately, then $PAC$ can be stated as $(k \times n) + ROW$ where $k$ is the number of such subsets. A higher $k$ implies higher losses in throughput. With increasing $PAC$ costs, there is a lower impetus for processing using PIM (instead of processing on the CPU) unless the cost is amortized over time. Additionally, the trade-offs may shift with the number of rows in a memory array and with any additional hardware support available for fast relocation (\textit{e.g.}, parallel vertical moves). Finally, while Fig.~\ref{fig:homo} showed CPU vs. PIM trade-offs assuming PAC is zero, a higher PAC too, just like OC, will adversely impact PIM's performance.}

\textbf{Energy Efficiency Impact.} As shown in Fig.~\ref{fig:power}, and based on Eq.~\ref{eq:3}, a maximum of {\bf 1950} $MATs$ can be accommodated for PIM  at the power envelope of 20W. Increasing the number of MATs does not further increase the throughput, since the power budget of the system is the main limiter. For example, at 40W up to 3900 memory arrays (MATs) can be active at any given time.  

For the CPU, the energy cost of data transfer limits the PL-Throughput-CPU. Here, we assume a $BW$ = 16Tbps. With a power limitation of 20W, the CPU delivers 55 GOPS at $DIO$ = 24. At a power budget of 40W, 111 GOPS are possible, and 444 GOPS at 160W. Compare this against the raw (with no power limitation) CPU throughput, which is 682 GOPS at a DIO of 24.

The values of the energy parameters $E^{PIM}$ and $E^{CPU}$ affect the relative energy efficiency of PIM versus the CPU. For example, consider the case of a single-bit NOR operation, where $OC$ = 1 (a single MAGIC operation) and DIO = 3 (2 input and 1 output bits). In this case, PIM consumes 1x$E^{PIM}$ = 0.1pJ while the CPU consumes 3x$E^{CPU}$ = 45pJ. For this example, the CPU energy consumption is approximately {\bf 450X} higher than that of PIM. However, as $OC$ increases, the relative efficiency of PIM decreases. For the limiting case of $OC$ = 7200 or higher (720pJ/0.1pJ = 7200), PIM becomes less attractive than the CPU with respect to energy efficiency. However, note that differences in energy parameter values will affect the relative merits of PIM or CPU. 

\newrevision{The above examples assume PAC = 0 for simplicity, but in reality a non-zero PAC will lead to higher energy consumption per operation, effectively reducing PIM's energy efficiency advantage over the CPU. We leave further analysis of these and other model parameters for future work.}

\section{Conclusions}
\label{sec:conclusion}

This paper motivates and describes Bitlet, an analytical model for PIM. We show how to use the model to find the cases in which PIM is beneficial and to understand the related trade-offs and limits, in a parameterized fashion. We hope the model will shed more light on the new bitwise-PIM paradigm.  


\section{Acknowledgement}

This work was supported by the European Research Council through the European Union’s Horizon 2020 Research and Innovation Programme under Grant 757259 and by the Israel Science Foundation under Grant 1514/17.

\end{document}